\documentclass[11pt]{article}

\usepackage{tikz}
\usetikzlibrary{arrows.meta,positioning,fit}

\usepackage[margin=1in]{geometry}
\usepackage{setspace}
\usepackage[T1]{fontenc}
\usepackage{lmodern}

\usepackage{amssymb}
\usepackage{amsmath}
\usepackage{array}

\usepackage{tabularx}
\usepackage{booktabs}
\usepackage{graphicx}
\usepackage{subcaption}

\usepackage{algorithm}
\usepackage{algorithmic}
\usepackage[hidelinks]{hyperref}

\usepackage[style=apa,natbib=true]{biblatex}
\let\cite\citep

\usepackage[hidelinks]{hyperref}
\usepackage{orcidlink}

\title{Evolutionary Systems Thinking:\\
From Equilibrium Models to Open-Ended Adaptive Dynamics}

\author{
Dan Adler \orcidlink{0009-0007-7122-0317} \\
\texttt{dan@danadler.com} \\
\small Presented at the 44th \textbf{International System Dynamics Conference} (ISDC) , July 23, 2026
}

\date{}  

\begin{document}

\maketitle


\begin{abstract}
Complex change is often described as ``evolutionary'' in economics, policy,
technology, and organizations, yet many system dynamics models represent
behavior within a fixed set of stocks, flows, relationships, and transition rules.
Such models can generate nonlinear, oscillatory, path-dependent, or chaotic
behavior, but structural novelty must ordinarily be specified in advance.
This paper argues that evolutionary dynamics should be treated as a core
systems-thinking problem rather than as a biological metaphor.

We introduce Stability-Driven Assembly (SDA), a minimal non-equilibrium
framework in which stochastic interactions and differential persistence
generate endogenous selection without genes, template-based replication, or
an externally specified fitness function. Longer-lived configurations
accumulate in the population and therefore become more likely to participate
in subsequent interactions. This creates feedback among persistence,
population composition, and future pattern formation. The resulting
abundance-weighted sampling is equivalent to fitness-proportional selection,
allowing SDA to be interpreted as a natural genetic algorithm driven by
persistence-weighted population dynamics.

SDA provides a conceptual basis for distinguishing fixed-state-space dynamics from
evolving possibility spaces, in which persistent structures can reshape
future flows, interactions, and available configurations. It also suggests
that equilibrium should be treated as provisional: a quasi-stable regime may
be reorganized when a more persistent configuration emerges. We conclude by
outlining two ways to extend system dynamics practice: constructing an
SDA-style population model alongside a stock--flow model, and using SDA
perturbation analysis to examine the vulnerability of an existing regime to
structural innovation.

\end{abstract}

\section{Introduction}

The term \emph{evolution} is widely used in economic, policy,
technological, and organizational studies to describe long-term change.
System dynamics provides a mature framework for representing feedback,
accumulation, delays, and nonlinear behavior
\cite{sterman2000business,meadows2008thinking}. Yet in many system
dynamics models, the relevant stocks, flows, relationships, and transition
rules are specified before the simulation begins. Such models
can exhibit nonlinear behavior, oscillation, path dependence, multiple
attractors, and collapse, but they ordinarily describe change within a
predefined structure. New strategies, institutions, technologies, or
categories can appear only if the modeler has already represented them
as possible states or transitions. This gap has long motivated evolutionary 
approaches in economics \cite{nelson1982evolutionary}, but it remains difficult to represent the endogenous construction of new categories and interaction structures.

The central distinction is between a system that moves through a fixed
possibility space and one that changes the space through which it can
subsequently move. In a conventional dynamical model, the state
variables may change, while the identities of those variables and the
rules governing their interaction remain fixed. In an evolutionary
system, by contrast, persistent structures generated during the process
can become part of the causal environment for future dynamics. A
strategy can create a market niche, an organizational routine can enable
new capabilities, and a policy can create agencies, constituencies, and
constraints that shape later policy choices. System dynamics makes
behavior endogenous to feedback structure; evolutionary system dynamics
must additionally allow aspects of that structure to become endogenous
to accumulated history.

This distinction reflects a broader issue in scientific explanation.
Idealized models \cite{cartwright1983laws} often
succeed by suppressing history, population heterogeneity, and contextual
variation in order to isolate tractable causal relationships. As
philosophers of science have emphasized, idealization is not a flaw but a
necessary methodological tool \cite{potochnik2017idealization}. However,
when the phenomenon of interest is inherently historical, cumulative,
and population-dependent, fixing the relevant entities and transition
structure in advance can remove the mechanism that requires explanation.
In such cases, simulation-based approaches may offer explanatory value
not simply because they contain more detail, but because they preserve
feedback among system state, population composition, and the formation
of future structure.

This distinction is developed more fully in
\cite{adler2026constructive} as a contrast between \emph{episodic} and
\emph{constructive} epistemic regimes. Episodic models treat successive
interactions as applications of invariant rules within a fixed state
space, whereas constructive systems allow the products of earlier
interactions to persist, accumulate, and become causal conditions for
later ones. The present paper translates that philosophical distinction
into a system dynamics context by asking how population-level memory can
make aspects of feedback structure endogenous to the history of the
system.

Darwinian evolution is conventionally modeled through heritable
variation, with heredity usually implemented by molecular or
template-based replication \cite{fisher1930genetical}. Before such
machinery existed, however, physical and chemical configurations already
differed greatly in how long they persisted under prevailing conditions.
Across physics, chemistry, and materials science, differences in
persistence arise from energetic minima, kinetic barriers, symmetry,
spatial organization, and hierarchical assembly, spanning timescales
from transient states to structures that endure for geological
durations. This raises a more general question: can differential
persistence create population-level selection before high-fidelity
replication exists?

In an open system with continued stochastic formation and turnover,
persistence can bias population composition. At comparable formation
rates, configurations that last longer accumulate more instances and
remain available for more interactions. Their increased abundance then
raises the probability that they will participate in subsequent
formation events. This closes a feedback loop:
persistence shapes population composition, population composition biases
future interactions, and those interactions generate configurations
whose persistence further reshapes the population. Selection therefore
need not be imposed as an external criterion; it can emerge from the
coupling of differential persistence, recurrent formation, and
abundance-weighted interaction.

In this paper, we use \emph{Stability-Driven Assembly} (SDA)
\cite{adler_sda} as a minimal framework to make this feedback
explicit. SDA combines stochastic interaction, finite lifetimes,
continuous replenishment, and removal of expired configurations in an
open, non-equilibrium population. Longer-lived configurations become
overrepresented and are consequently sampled more often in later
interactions. This produces abundance-proportional sampling equivalent
to fitness-proportional selection when persistence acts as an implicit
fitness measure. With recombination and mutation, the same mechanism can
be interpreted as a natural genetic algorithm, SDA/GA
\cite{adler2026sda_ga}, without an externally specified objective
function or template-based replication.

The purpose of the paper is not to argue that conventional system
dynamics models must converge to equilibrium, nor that agent-based
simulation is intrinsically evolutionary. Rather, it is to identify the
additional mechanism required when persistent structures alter future
flows, interactions, or available configurations. We first distinguish
fixed-state-space dynamics from evolving possibility spaces and present the
minimal SDA mechanism and illustrative simulation results. We then show
how differential persistence can produce quasi-stable regimes that are
punctuated when a more persistent configuration emerges and reorganizes
population flows. Finally, we discuss the relationship of SDA to
stock--flow and agent-based modeling and propose two practical
extensions to system dynamics: constructing an SDA-style population
model alongside an existing stock--flow model, and using SDA
perturbation analysis to examine the vulnerability of an established
regime to structural innovation.

\section{Fixed-state-space Dynamics and Evolving Possibility Spaces}
\label{sec:state_spaces}

The central distinction developed in this paper is not between systems
that reach equilibrium and systems that remain in motion. A conventional
system dynamics model may exhibit oscillation, overshoot, path dependence,
multiple attractors, chaos, or collapse while remaining structurally fixed.
The relevant distinction is, instead, between dynamics within a predefined
possibility space and dynamics that can modify the effective space of
future possibilities.

A conventional dynamical system may be represented schematically as
\begin{equation}
    \frac{dx}{dt} = F(x;\theta),
    \label{eq:fixed_dynamics}
\end{equation}
where the state variables $x$, the parameters $\theta$, and the functional
form of the interactions are specified before the simulation begins.
The values of the variables change over time, but the identities of the
variables and the repertoire of possible transitions remain fixed. 
Therefore, any new strategy, institution, technology, or organizational form must
be represented in advance, even if its eventual occurrence is uncertain.

An evolutionary system requires an additional form of endogeneity. Let
$P(t)$ denote the population of persistent configurations generated by
the system's history. The effective dynamics may then be represented
schematically as
\begin{equation}
    \frac{dx}{dt} = F\bigl(x;P(t)\bigr),
    \label{eq:evolving_dynamics}
\end{equation}
where the accumulated population structure modifies future flows,
interaction probabilities, or accessible transitions. The notation is
schematic rather than a complete formalism: its purpose is to express
that the effective dynamics depend on structures produced during the
process itself.

This distinction can be summarized as follows. A fixed-state-space dynamical
system moves through a possibility space specified by the modeler. An
evolutionary system can construct persistent configurations that become
part of the causal environment and thereby alter what can occur next.
System dynamics makes behavior endogenous to feedback structure;
evolutionary system dynamics additionally asks how aspects of that
structure can become endogenous to accumulated history.

\begin{figure}[h]
    \centering
    \includegraphics[width=1\textwidth]
    {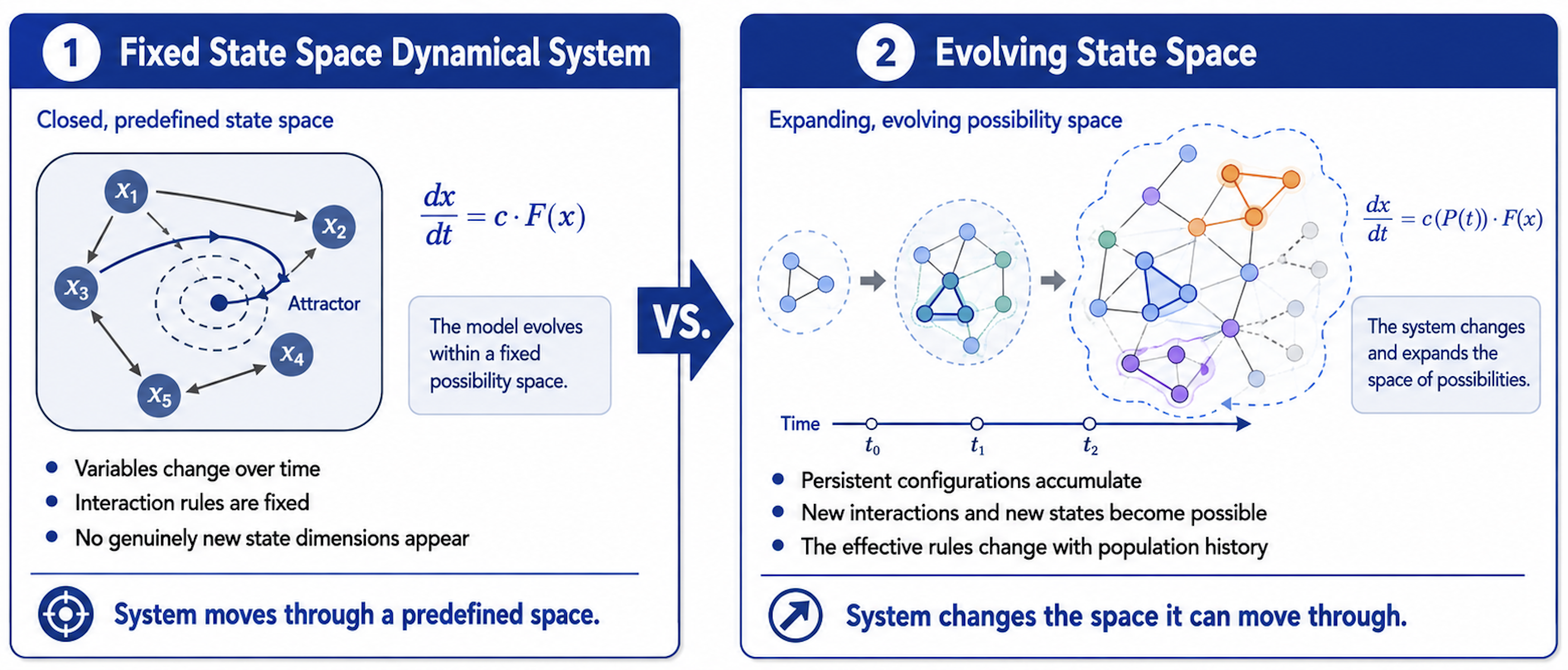}
    \caption{Fixed-state-space models move through a predefined possibility
    space, whereas evolutionary systems can reshape the effective space
    of future possibilities as persistent structures accumulate.}
    \label{fig:fixed_evolving}
\end{figure}

In economic, technological, organizational, and policy systems, such
persistent configurations can include strategies, institutions,
standards, infrastructures, routines, metrics, or regulatory
arrangements. A persistent strategy may create niches for complementary
strategies; an organizational routine may enable new capabilities; and a
policy may create agencies, constituencies, and constraints that shape
future policy choices. These structures do not merely change the current
state of the system. They modify the conditions under which subsequent
states are generated.

Evolutionary language is therefore mechanistically meaningful only when
the model identifies how variants arise, why some persist longer than
others, and how the resulting population changes subsequent dynamics.
Competition, learning, imitation, and adaptation may contribute to this
process, but they do not by themselves guarantee an evolving possibility
space. The decisive condition is that the structures produced by earlier
interactions persist and feed back into the generation of later ones.

This criterion also clarifies the relationship between evolutionary
dynamics and evolutionary game theory. Standard evolutionary game
models describe changes in the frequencies of strategies within a
specified set of strategies and payoff structure \cite{nowak2006evolutionary}.
Such models are evolutionary in the population-dynamical sense, but they
do not necessarily represent the endogenous construction of new
strategies, payoff relations, or categories. The present paper addresses
this additional constructive dimension: how population history can
reshape the effective structure of the system itself.

\section{Stability-Driven Assembly as an Evolutionary Mechanism}
\label{sec:sda}

Stability-Driven Assembly (SDA) is a minimal non-equilibrium framework
in which differential persistence biases the accumulation of structure
over time \cite{adler_sda}. A companion study extends the framework
through recombination and mutation and applies it to chemical symbol
space, where it produces scaffold-level dominance, entropy reduction,
and sustained novelty \cite{adler2026sda_ga}. The term ``assembly''
here refers to pattern formation through interaction and is distinct
from Assembly Theory \cite{sharma2023}, which measures constructional
complexity rather than modeling population-level selection.

The core SDA mechanism is intentionally simple. Patterns are generated
through stochastic interactions in an open population, assigned finite
lifetimes, and removed when those lifetimes expire. Base elements are
continually replenished to maintain turnover and continued exploration.
Relative to unconstrained random assembly, the decisive addition is
differential persistence.

\begin{figure}[htbp]
    \centering
    \includegraphics[width=0.9\textwidth]{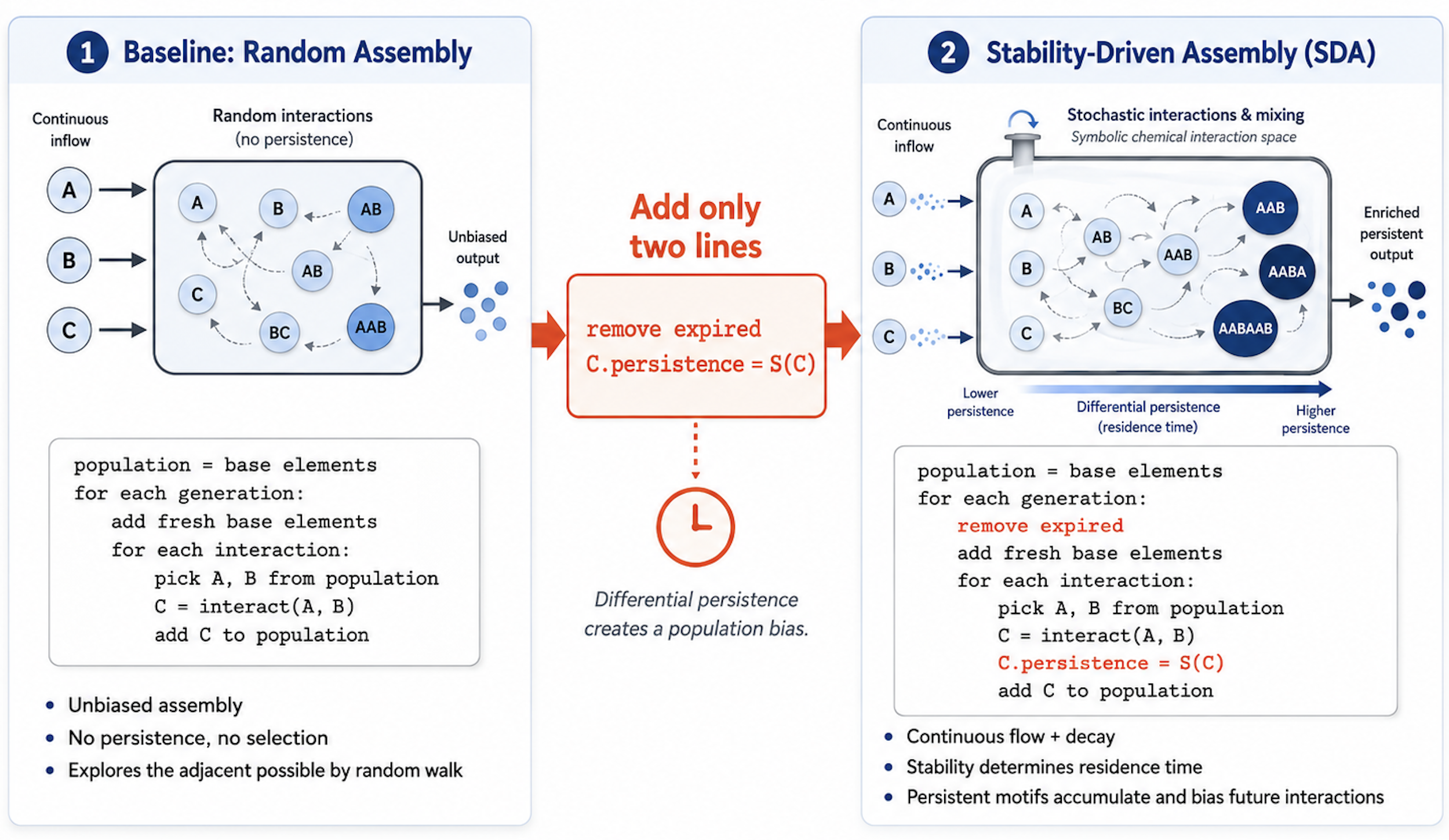}
    \caption{From random assembly to Stability-Driven Assembly.
    In the baseline case, stochastic interaction alone performs an
    unbiased search over nearby possibilities. Adding only two
    operations---removal of expired patterns and assignment of
    persistence according to a stability function $S$---creates a
    population bias. Persistent motifs remain in the population longer,
    accumulate, and participate more often in subsequent interactions.}
    \label{fig:random-to-sda}
\end{figure}

Figure~\ref{fig:random-to-sda} provides a conceptual comparison with
purely accumulative random assembly. The simulation in
Section~\ref{sec:sda-result} uses a stricter uniform-persistence control:
both conditions include expiration and differ only in whether lifetimes
vary among patterns.

\begin{figure}[htbp]
    \centering
    \includegraphics[width=0.5\textwidth]{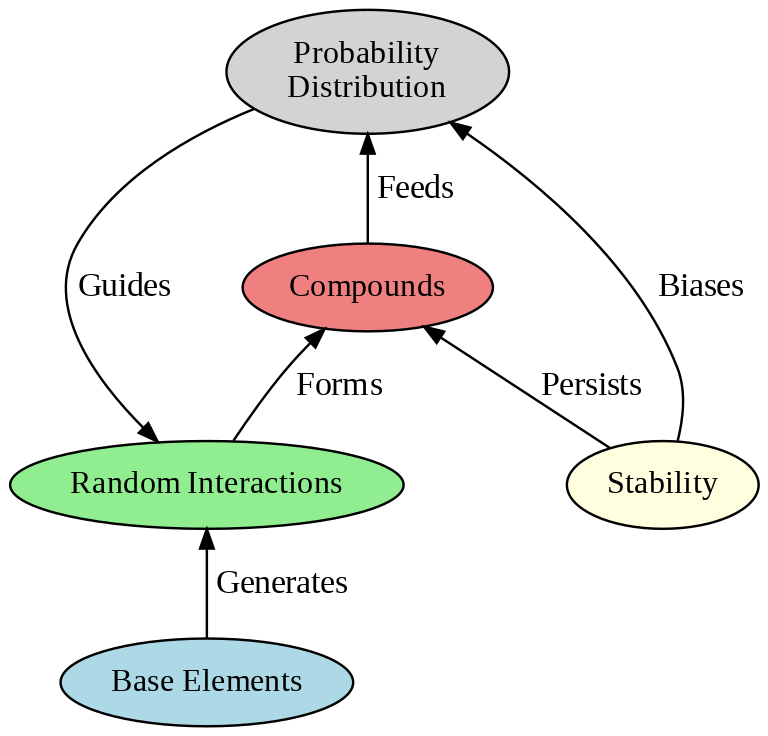}
    \caption{Core feedback structure underlying SDA. Stochastic
    interactions generate patterns; differential persistence reshapes
    population composition; and the evolving population distribution
    biases future interactions. Selection emerges endogenously from this
    feedback loop.}
    \label{fig:sda-feedback}
\end{figure}

Patterns that persist longer remain available for more interactions and,
when recurrently generated, accumulate more instances. Their greater
abundance raises their probability of participating in subsequent
interactions. This closes the central SDA feedback loop shown in
Figure~\ref{fig:sda-feedback}.

This mechanism produces selection without an externally imposed
selection operator. Stochastic interaction supplies variation, while
differential persistence alters the population from which subsequent
interactions are sampled. When the interaction operator includes
recombination and mutation, abundance-proportional sampling plays the
role of fitness-proportional or roulette-wheel selection
\cite{holland1975adaptation,goldberg1989genetic}. SDA can therefore be
interpreted as a natural genetic algorithm, SDA/GA
\cite{adler2026sda_ga}, in which persistence acts as an implicit
fitness measure.

The analogy does not require biological lineage or template copying.
Multiple instances of a motif may arise through independent interaction
events and share structural identity without descending from a common
replicating individual. More generally, persistence should be understood
as a context-dependent residence time. In physical systems, it may depend
on bonding, activation barriers, or spatial organization; in
socio-technical systems, it may depend on switching costs, legitimacy,
network effects, complementary capabilities, or institutional support.
In each case, the population-level consequence is the same:
configurations that remain active longer have more opportunities to
participate in and influence subsequent dynamics.

\subsection{Illustrative Symbolic Result}
\label{sec:sda-result}

To illustrate the population-level effect of differential persistence,
we compare a minimal symbolic SDA simulation with a
uniform-persistence control. In both cases, patterns are generated
stochastically through interaction, expired instances are removed, and
base elements are continuously replenished. In the SDA condition,
lifetimes vary according to the stability function $S(p)$; in the
control condition, all patterns receive the same lifetime. No explicit
fitness function or selection operator is applied
\cite{adler_sda,adler2026sda_ga}.

\begin{figure}[h]
\centering
\includegraphics[width=0.6\textwidth]{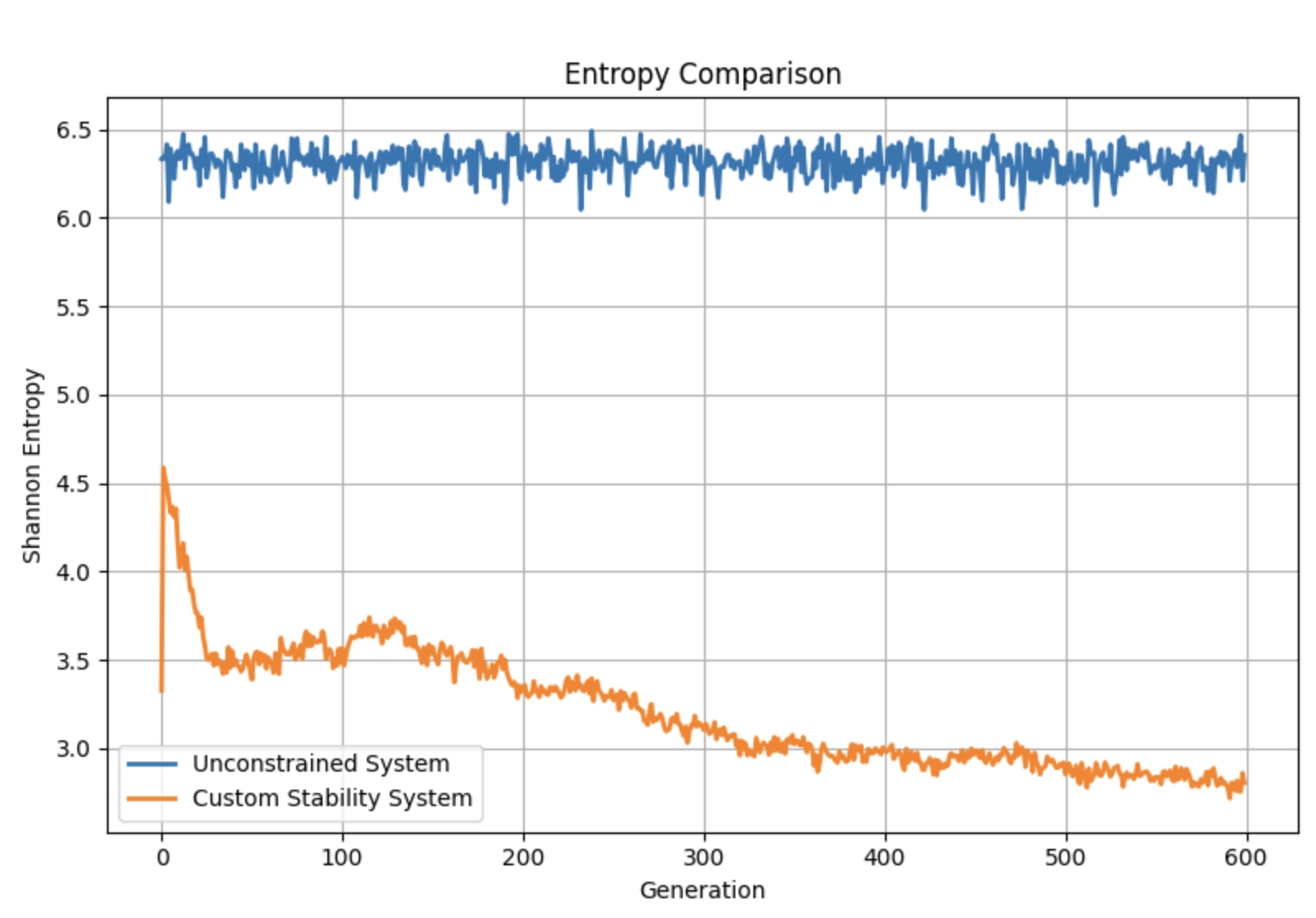}
\caption{Shannon entropy of the population distribution under SDA and
a uniform-persistence control. Entropy remains comparatively high when
all patterns receive the same lifetime. Under differential persistence,
probability mass becomes concentrated on a smaller set of long-lived
motifs, producing lower population entropy.}
\label{fig:sda-entropy}
\end{figure}

Figure~\ref{fig:sda-entropy} shows the Shannon entropy of the
population distribution over time. In the uniform-persistence control,
continued stochastic formation maintains a relatively dispersed
distribution. Under SDA dynamics, entropy declines as longer-lived
motifs accumulate and become more frequently represented in subsequent
interactions. The resulting concentration of probability mass
demonstrates that differential persistence, coupled to recurrent
formation and abundance-proportional sampling, is sufficient to
generate endogenous selection pressure without an explicit fitness
function or template-based replication. This establishes the minimal selection mechanism used in the remainder of the paper. It does not, by itself, demonstrate open-ended
evolution or indefinite state-space expansion; rather, it shows how
persistence-weighted population feedback can transform stochastic
assembly into directed, path-dependent search. Extended simulations,
intervention experiments, and chemical-symbol applications are reported
elsewhere \cite{adler_sda,adler2026sda_ga}.

\section{Population Memory and Structural Change}
\label{sec:population_memory}

The symbolic result above establishes a minimal population-level
mechanism: differential persistence changes abundance, and abundance
biases the interactions from which later configurations are generated.
The population therefore retains information about earlier formation
events. History is represented not only by the current values of
aggregate variables, but also by the configurations that remain active
and continue to participate in the system.

Important precursors to this perspective emphasized the historical and
non-equilibrium character of evolutionary processes. Brooks and Wiley's
\emph{Evolution as Entropy} \cite{brooks1986evolution} rejected
teleological and optimization-based accounts in favor of irreversibility,
constraint, and information accumulation. Related work in econophysics
challenged equilibrium and representative-agent assumptions by modeling
economies as heterogeneous populations of interacting entities
\cite{farmer2005economics,farmer2009economy}. SDA contributes a specific
mechanism through which such history becomes causally active:
persistent configurations remain available for reuse and thereby alter
the population from which subsequent interactions are sampled.

However, persistence-driven selection and structural evolution are not the same. Differential persistence alone can
concentrate a population on a subset of configurations, as shown in
Section~\ref{sec:sda-result}. A stronger evolutionary effect occurs when
those accumulated configurations also modify later formation rates,
interaction opportunities, persistence conditions, or aggregate flows.
Population memory then changes not only which configurations are common,
but also the effective conditions under which future configurations are
generated and retained.

It is also useful to distinguish SDA from standard birth--death and
predator--prey models. Such models already represent turnover and
population feedback: births and deaths change population abundance, and
the abundance of one population may affect the growth or decline of
another \cite{kendall1948generalized}. In their standard formulations,
however, the relevant population types and interaction channels are
specified in advance. SDA does not differ merely because it includes
decay or population-dependent interaction. It adds a constructive layer
in which interactions generate configurations, differential persistence
determines which remain available, and the resulting population biases
the formation of subsequent configurations. The distinction is therefore
between feedback among predefined populations and feedback through which
the active population repertoire can itself be constructed and reshaped.
Models with evolving traits, adaptive interactions, or endogenous network
structure may cross this boundary and are best understood as neighboring
approaches to evolutionary system dynamics.

The present symbolic model demonstrates persistence-driven selection
and provides a basis for investigating this stronger structural effect.
It does not by itself establish unlimited novelty or literal expansion
of state-space dimensionality. Its more specific contribution is to
identify persistence-weighted population feedback as a bridge from
stochastic formation to path-dependent structural change. The following
sections consider how this mechanism can be represented using
agent-based and stock--flow models.

\section{Agent-Based Modeling and Evolutionary System Dynamics}
\label{sec:abm}

\subsection{ABM as an Implementation Substrate}

Agent-based models (ABMs)
\cite{hamill2016agent,resnick1994turtles,railsback2012agent}
provide a natural computational substrate for evolutionary system
dynamics because they can represent heterogeneous entities, stochastic
interaction, entry and exit, local adaptation, and endogenous network
formation. Related approaches in evolutionary economics likewise model
economic change through populations of firms, routines, technologies,
and institutions rather than through representative agents or static
equilibrium alone \cite{dopfer2008,hodgson2010}.

Agent-based representation, however, does not by itself make a model
evolutionary in the constructive sense developed here. An ABM may
contain heterogeneous and adaptive agents while retaining a predefined
set of agent types, strategies, interaction rules, and possible
transitions. Such a model can produce complex emergent behavior within
its specified architecture without allowing structures generated during
the simulation to modify the effective space of subsequent possibilities.

Conversely, ABMs are not intrinsically limited to fixed structures.
They can be designed so that strategies, networks, organizations, or
interaction rules emerge and subsequently alter agent behavior. The
relevant question is therefore not whether a model is agent-based, but
whether its generated population structure becomes causally active in
future dynamics.

SDA supplies one minimal mechanism for creating this coupling.
Differential persistence changes the abundance of configurations in the
population, and abundance changes the probability that those
configurations participate in subsequent interactions. The population
distribution is therefore not merely a summary statistic. It is part of
the mechanism determining which interactions occur and which new
configurations become accessible.

This distinction may be expressed as a design criterion for evolutionary
ABMs. Such a model should represent the generation of structural
variation; differential persistence, removal, or turnover; interaction
probabilities that depend on the resulting population composition; and
feedback through which persistent structures modify later flows,
interactions, or available configurations. An ABM satisfying these
conditions could implement SDA directly or realize a related evolutionary
mechanism. SDA should therefore be understood not as an alternative to
agent-based modeling, but as a population-feedback principle that
agent-based simulation is especially well suited to represent.

\subsection{Stochastic Interaction and Differential Persistence}

In many physical, social, and organizational systems, novelty originates
through contingent interaction. Molecules encounter one another,
individuals form collaborations, firms establish partnerships, and
institutions combine existing procedures into new organizational forms.
Most such encounters leave no durable trace, while a smaller number
produce configurations that remain available for further interaction.

Persistence acts as a population-level filter on this variation.
A durable collaboration may attract additional participants; a
successful routine may become embedded in organizational practice; and
a technical standard may acquire complementary products and
infrastructure. These structures need not copy themselves directly.
Because they remain present, they receive more opportunities to
participate in subsequent events and to contribute components to later
configurations.

This is the mechanism formalized by SDA. Stochastic interaction supplies
exploration, while differential persistence changes the population from
which later interactions are sampled. Selection arises through unequal
continued participation rather than through centralized choice or an
externally specified optimization criterion. Persistent structures need
not be globally optimal; they need only remain sufficiently viable under
prevailing conditions to accumulate and influence subsequent dynamics.

Stochasticity is useful because it continually exposes the system to new
combinations, but randomness alone is not sufficient for evolutionary
change. Without persistence, accumulation, and population feedback,
novel configurations disappear without affecting what happens next.
Likewise, persistence without continued variation may produce
concentration without sustained innovation. Evolutionary dynamics arise
from the coupling of exploration with differential retention and reuse.

\section{Adding Evolutionary Dynamics to Stock--Flow Models}
\label{sec:adding_evolution}

One of the enduring strengths of system dynamics is the unifying power
of the stock--flow formalism
\cite{sterman2000business,meadows2008thinking,garcia2019theory}.
Inventories, bank balances, populations, chemical concentrations, and
other accumulated quantities can be represented using a common language
of stocks, inflows, outflows, delays, and feedback. This abstraction
allows behavior to emerge endogenously from a model's feedback
structure.

Evolutionary systems require an additional form of endogeneity. The
issue is not merely that stock values change, but that persistent
structures generated during the process may alter the flows,
relationships, and possibilities available afterward. A new strategy can
create opportunities for complementary strategies; an organizational
routine can enable previously unavailable capabilities; and a policy can
create agencies, constituencies, metrics, and constraints that influence
later interventions. The products of the dynamics become part of the
causal environment for subsequent dynamics.

In an SDA representation, stocks correspond to populations of
configurations rather than only to quantities of predefined substances.
Formation events create new configurations, while expiration and removal
provide their outflows. Differential persistence changes the residence
time of each configuration, and the resulting abundance distribution
biases later interaction flows. New configurations may therefore enter
the active population, become causally important, and eventually decline
or disappear.

This does not require abandoning stock--flow modeling. Rather, it
requires coupling a conventional aggregate model to a population-level
constructive process. The stock--flow model represents the current
macroscopic regime, while the SDA component represents the generation,
persistence, and selection of structures capable of modifying that
regime.

\subsection{Quasi-Stable Regimes and Punctuated Reorganization}
\label{sec:punctuated}

Differential persistence can produce periods in which a small number of
configurations dominate the population and the associated flows become
approximately stable. Such a regime may resemble equilibrium from the
perspective of an aggregate stock--flow model, even though it is
maintained by continuing formation, expiration, and replacement at the
population level. We refer to this as a \emph{quasi-stable regime}.

Quasi-stability is provisional. Suppose that configurations $A$ and $B$
have accumulated substantial population mass and jointly support the
current pattern of interactions. If a new configuration $C$ is generated
or introduced and persists longer under the prevailing conditions, it
may accumulate, participate increasingly in subsequent interactions,
and redirect population flows away from the previously dominant
structures. The system may then reorganize around a new dominant
configuration.

\begin{figure}[h]
    \centering
    \includegraphics[width=1\textwidth]
    {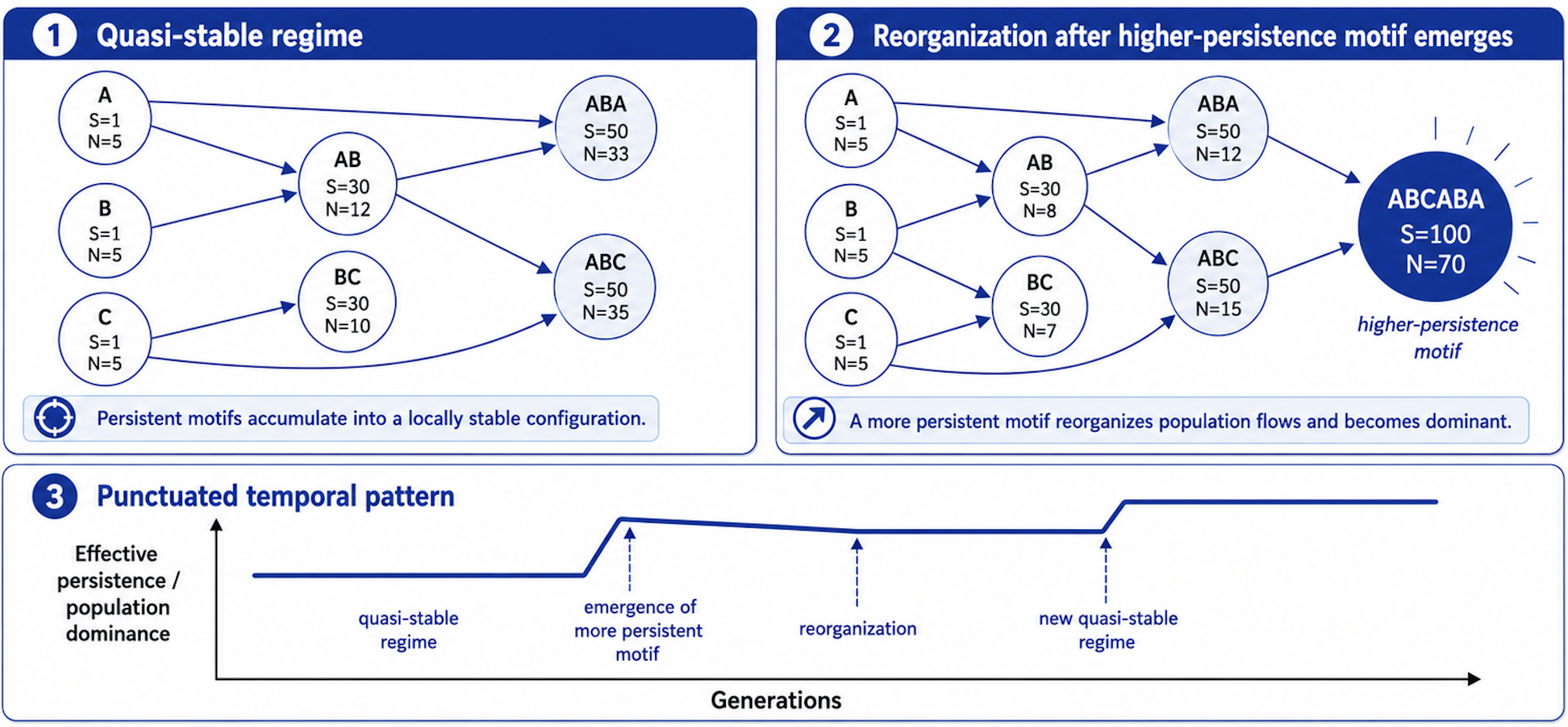}
    \caption{Punctuated reorganization in an SDA population. A
    quasi-stable regime dominated by persistent configurations can be
    disrupted when a higher-persistence configuration emerges or is
    introduced. As the new configuration accumulates, it redirects
    interaction flows and establishes a new quasi-stable regime.}
    \label{fig:punctuated-reorganization}
\end{figure}

Figure~\ref{fig:punctuated-reorganization} illustrates the resulting
temporal pattern: extended periods of apparent stability interrupted by
comparatively rapid reorganization. Similar patterns are described as
punctuated equilibrium in evolutionary biology, creative destruction in
economics, regime change in policy studies, and technological disruption
in innovation research. SDA does not claim that these phenomena share
identical domain-specific mechanisms. It identifies a common
population-level structure: persistent configurations concentrate
interaction opportunities until the emergence of a sufficiently
persistent alternative reorganizes those opportunities.

This distinction also clarifies what an evolutionary extension may add
to an existing system dynamics model. If collapse, instability, or
regime transition is already represented as an attractor within the
model's specified structure, SDA need not improve its prediction. SDA
becomes relevant when the transition depends on a new strategy,
technology, institution, organizational form, or interaction structure
that is not represented in the original state space.

An equilibrium identified by a stock--flow model may therefore be
locally valid while remaining vulnerable to structural innovations
outside the model's specified repertoire. The equilibrium is not
necessarily wrong; it may be incomplete as a long-term description.

\subsection{Illustrative Domain Mappings}
\label{sec:illustrative-domains}

In system dynamics applications, persistent configurations are generally
not molecules or symbolic strings. They are domain-specific structures
such as strategies, routines, standards, technologies, institutions, and
policy arrangements. Although the mechanisms responsible for their
persistence differ across domains, they can play the same dynamical role:
once established, they remain available for further interaction and
become part of the environment shaping what can occur next.

\begin{figure}[htbp]
    \centering
    \includegraphics[width=1\textwidth]
    {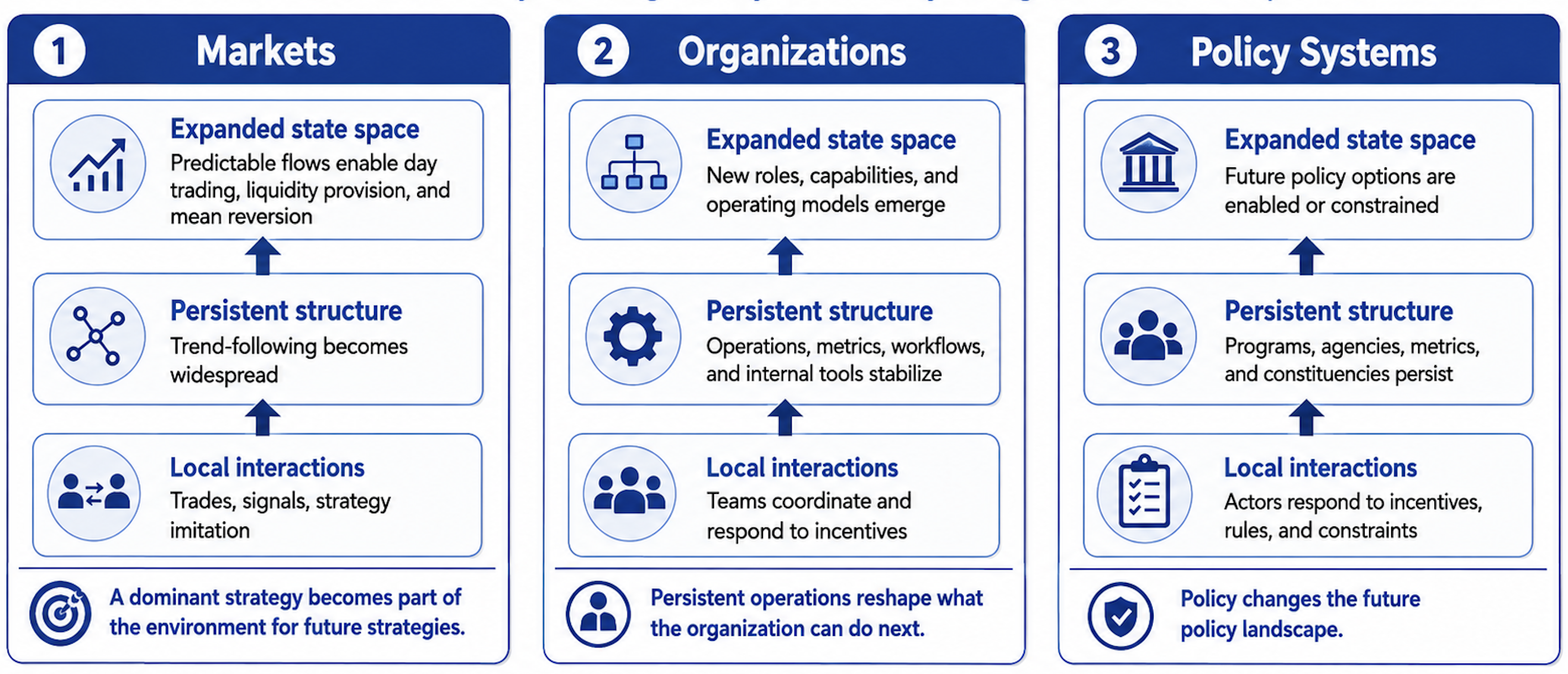}
    \caption{Illustrative pathways through which persistent structures
    can reshape future possibilities in three system dynamics domains.
    In markets, a widely adopted strategy can create niches for
    complementary strategies and services. In organizations, persistent
    operations, metrics, workflows, and tools can enable new roles and
    capabilities. In policy systems, programs, agencies, metrics, and
    constituencies created by one intervention can enable or constrain
    subsequent policy choices. These examples are conceptual mappings
    rather than calibrated models.}
    \label{fig:evolving-domains}
\end{figure}

\paragraph{Markets.}
Local trades, signals, and imitation can cause particular strategies to
become widespread. Once a strategy such as trend-following becomes a
persistent feature of the market, it changes the environment faced by
other participants. It may create opportunities for liquidity provision,
mean-reversion strategies, execution services, risk-management products,
or regulatory responses. The persistent strategy therefore does more
than increase its own frequency: it helps alter the effective strategy
space available to others.

\paragraph{Organizations.}
Teams continually develop routines, metrics, workflows, and internal
tools in response to local problems and incentives. Some disappear when
the immediate problem passes, while others become embedded in
organizational practice. Once persistent, these structures can support
new roles, capabilities, reporting relationships, and operating models.
Organizational history is therefore retained not only in aggregate
performance measures, but in the operational structures that determine
what the organization can subsequently do.

\paragraph{Policy systems.}
Policy interventions frequently create structures that outlast the
initial decision: programs, administrative agencies, reporting
requirements, metrics, funding streams, legal precedents, and organized
constituencies. These persistent products of earlier policy become
constraints and affordances for later interventions. Policy therefore
changes not only the current state of the system, but also the policy
landscape within which future decisions are made.

Across all three domains, the common mechanism is a progression from
local interaction to persistent structure to altered future
possibilities. The relevant persistence function is context-dependent:
a strategy, routine, or institution remains active because of factors
such as profitability, switching costs, legitimacy, complementary
resources, legal authority, network effects, or organizational memory.
SDA abstracts these different mechanisms into their shared
population-level consequence: longer-lived structures have more
opportunities to participate in and influence subsequent dynamics.

These examples do not establish that every persistent structure produces
an evolving possibility space. The additional condition is that the
structure must affect later formation, interaction, or persistence.
Persistence that merely preserves an object is retention; persistence
that changes what can form or remain viable next becomes evolutionary
feedback.

\subsection{Two Approaches to Evolutionary Extension}
\label{sec:two-strategies}

The SDA perspective suggests two complementary ways to extend system
dynamics practice. The first introduces evolutionary dynamics directly
by coupling a population-level constructive model to a stock--flow
model. The second retains an existing system dynamics model and uses
SDA-style perturbations to examine whether a reference regime is
vulnerable to structural innovations not represented in the original
model.

\begin{figure}[htbp]
    \centering
    \includegraphics[width=0.98\textwidth]
    {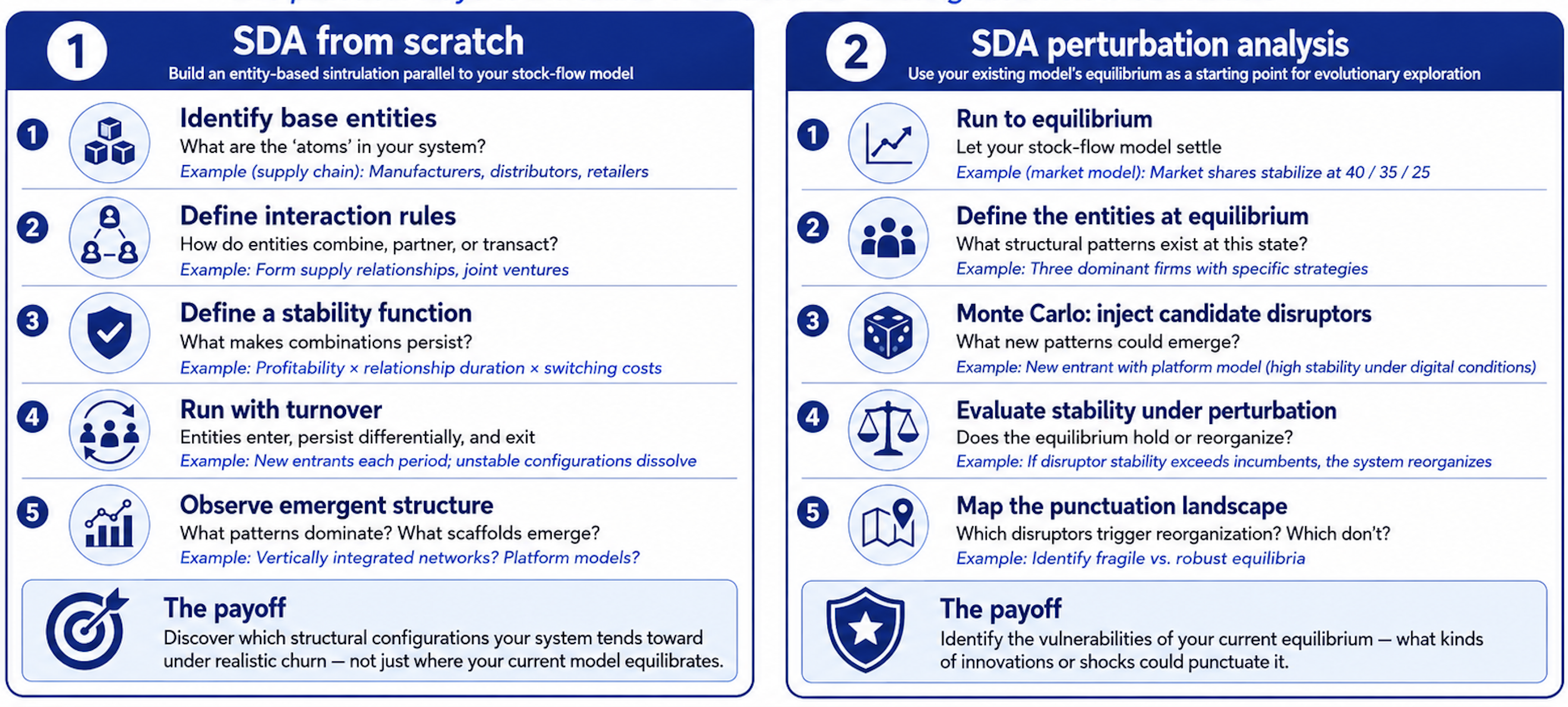}
    \caption{Two proposed approaches to introducing evolutionary
    dynamics into system dynamics models.
    \textbf{Left:} A coupled SDA population model represents the
    formation, context-dependent persistence, turnover, and accumulation
    of entities such as strategies, technologies, institutions, or
    organizational configurations.
    \textbf{Right:} SDA perturbation analysis begins from a reference
    regime generated by an existing stock--flow model and introduces
    candidate structural disruptors to test whether they disappear,
    coexist with existing structures, or reorganize the regime.}
    \label{fig:sda-two-strategies}
\end{figure}

The two strategies answer different questions. A coupled SDA population
model asks which structures emerge and become dominant under continuing
variation and turnover. SDA perturbation analysis asks a conditional
question: if a configuration with specified persistence and coupling
properties appeared, would the existing regime absorb it or reorganize?

\section{Toward Evolutionary System Dynamics}
\label{sec:toward_esd}

The preceding analysis suggests three progressively stronger forms of
endogeneity. In conventional system dynamics, behavior is endogenous to
a feedback structure specified by the modeler. In population-based
evolutionary models, the relative abundance of competing configurations
also changes endogenously through formation, interaction, persistence,
and turnover. In a constructive evolutionary model, persistent
configurations additionally modify the effective interactions,
persistence conditions, or structural possibilities that shape later
dynamics.

Evolutionary system dynamics is concerned primarily with this third
form. It seeks to couple population-level construction and selection to
aggregate stock--flow behavior so that structures generated during the
process can become part of the causal environment for subsequent
change. The aim is not to abandon conventional system dynamics, but to
extend it from the endogenous evolution of system behavior toward the
endogenous evolution of aspects of the effective feedback structure
itself.

SDA provides a minimal mechanism for connecting the second and third
levels. Differential persistence changes population composition, and
population composition changes which configurations are most likely to
participate in later interactions. When persistent configurations also
modify aggregate flows or enable new combinations, population history
begins to reshape the effective possibility space of the system. The
result is not simply movement toward another predefined attractor, but
the endogenous construction of conditions that influence which
attractors or regimes can subsequently arise.

The present symbolic results demonstrate endogenous selection pressure
and population concentration; they do not by themselves establish
open-ended evolution in its strongest sense. A stronger demonstration
would require sustained generation of structural novelty, continued
turnover without convergence to a terminal repertoire, and feedback by
which generated structures modify the mechanisms of later variation,
interaction, or persistence. Novel configurations need not be listed
individually at initialization, although the model must still specify a
generative representation and a set of composition rules.

Agent-based modeling offers a useful implementation architecture for
such systems because it can represent heterogeneous configurations,
local interaction, entry and exit, endogenous networks, and
population-dependent behavior. SDA is not an alternative to agent-based
modeling, but a mechanism that can be implemented within it. The coupled
approach proposed in Section~\ref{sec:two-strategies} would use an
agent- or entity-based population model to generate and select
structures, while a stock--flow model represents the aggregate
environment that determines their persistence and is, in turn, modified
by their accumulation.

Artificial intelligence may expand the repertoire of generative
operators available within this architecture. For example, learning
agents or generative models could propose new strategies,
representations, or combinations that were not individually enumerated
by the modeler. AI does not, however, supply evolutionary dynamics by
itself. Unless generated variants undergo differential persistence and
feed back into the future generation process, learning merely searches
within an externally defined objective and representation. Its possible
role here is therefore as a source of structured variation, not as a
substitute for population-level selection.

The immediate research agenda is to test these ideas in established
system dynamics domains. This includes coupling SDA populations to
canonical stock--flow models, defining empirically meaningful
persistence functions, comparing SDA extensions with evolutionary
agent-based models, and evaluating whether perturbation analysis reveals
structural vulnerabilities that conventional parameter sensitivity
analysis does not. These studies would clarify when evolutionary
extensions add explanatory value and when a fixed-structure system
dynamics model remains sufficient.

\section{Conclusion}
\label{sec:conclusion}

This paper has argued that the central distinction between conventional
and evolutionary system dynamics is not whether a model reaches
equilibrium. System dynamics models can generate oscillation, collapse,
path dependence, multiple attractors, and other complex behavior while
remaining within a structure specified in advance. The relevant
distinction is between behavior that unfolds within a predefined
possibility space and dynamics in which persistent structures generated
during the process become part of the causal environment for what can
occur next.

Stability-Driven Assembly provides a minimal illustration of the
population feedback required for this additional form of endogeneity.
Stochastic interaction generates variation, differential persistence
changes population composition, and abundance-proportional sampling
biases subsequent interactions. Selection pressure therefore emerges
without an explicit selection operator, externally specified objective
function, or template-based replication. The symbolic results show that
this mechanism concentrates probability mass on persistent motifs and
transforms undirected assembly into path-dependent evolutionary search.

The present results do not by themselves establish open-ended evolution
in its strongest sense. They identify a more specific mechanism through
which population history can become causally active. When persistent
configurations also modify aggregate flows, interaction opportunities, or
the viability of later configurations, the system's effective possibility
space can change over time. Temporarily dominant structures may then
support quasi-stable regimes that are punctuated when a more persistent
configuration emerges or is introduced and reorganizes population flows.

For system dynamics practice, this perspective suggests two
complementary strategies. A coupled SDA population model can represent
the continuing formation, persistence, turnover, and selection of
strategies, technologies, institutions, or organizational forms alongside
an aggregate stock--flow model. Alternatively, SDA perturbation analysis
can begin from a regime generated by an existing model and test its
conditional vulnerability to candidate structural innovations. The first
models evolutionary construction directly; the second stress-tests a
specified regime for evolutionary fragility.

These approaches extend rather than replace conventional system
dynamics. Stock--flow models remain powerful tools for explaining
accumulation, delay, feedback, and behavior within a defined structure.
Evolutionary system dynamics adds the question of how structures
themselves arise, persist, alter later feedback, and are eventually
displaced. Agent-based modeling offers a useful implementation
architecture for this purpose, but the defining requirement is not the
presence of agents or complexity alone. It is feedback through which
generated population structure influences future formation and
persistence.

Evolution, on this account, is not merely a biological metaphor for
long-term change. It is a class of population-dependent dynamics in which
persistent products of history reshape the conditions of subsequent
change. Establishing when this extension adds explanatory value will
require applications to canonical system dynamics models, empirically
meaningful persistence functions, and comparisons with conventional
sensitivity and stress-testing methods. These provide a practical
research agenda for developing evolutionary system dynamics as a
complement to established systems modeling.

\printbibliography

\end{document}